\documentclass[letterpaper, 10 pt, conference]{ieeeconf}
\IEEEoverridecommandlockouts
\usepackage{multicol}
\usepackage[hyphens,spaces,obeyspaces]{url}
\usepackage{paralist}
\usepackage[ruled,vlined]{algorithm2e}

\everypar{\looseness=-1}
\usepackage{amsmath}
\usepackage{amssymb}
\usepackage{graphicx}
\usepackage{xcolor}
\usepackage{subcaption}

\newtheorem{defi}{Theorem}
\newtheorem{definition}[defi]{Definition}
\newtheorem{assu}{Assumption}

\newtheorem{rema}{Theorem}
\newtheorem{remark}[rema]{Remark}

\newtheorem{exam}{Theorem}
\newtheorem{example}[exam]{Example}
\newtheorem{prob}{Problem}

\newcommand{\sys}{\mathcal{E}}
\newcommand{\model}{\mathcal{M}}
\newcommand{\cont}{\pi}
\newcommand{\spec}{\varphi}

\newcommand{\vect}[1]{\mathbf{#1}}
\newcommand{\vu}{\vect{u}}

\newcommand{\vx}{\vect{x}}
\newcommand{\vy}{\vect{y}}

\newcommand{\vp}{\vect{p}}

\newcommand{\reals}{\mathbb{R}}
\newcommand{\integers}{\mathbb{N}}

\renewcommand{\baselinestretch}{0.97}

\begin{document}

\title{\LARGE \bf Counterexample-Guided Synthesis of Perception Models and Control}
\author{
Shromona Ghosh, Yash Vardhan Pant, Hadi Ravanbakhsh, and Sanjit A. Seshia \thanks{$^{\star}$ The second and fourth authors are with the EECS Department, UC Berkeley, CA 94720 USA. The first and third authors are with Waymo and Google respectively, but this work was done while they were at UC Berkeley. Correspondence to sseshia@berkeley.edu. This work is published at \textit{American Control Conference} 2021.}
}

\maketitle

\begin{abstract}
Recent advances in learning-based perception systems have led to drastic improvements in the performance of robotic systems like autonomous vehicles and surgical robots. These perception systems, however, are hard to analyze and errors in them can propagate to cause catastrophic failures. In this paper, we consider the problem of synthesizing safe and robust controllers for robotic systems which rely on complex perception modules for feedback. We propose a counterexample-guided synthesis framework that iteratively builds simple surrogate models of the complex perception module and enables us to find safe control policies. The framework uses a falsifier to  find counterexamples, or traces of the systems that violate a safety property, to extract information that enables efficient modeling of the perception modules and errors in it. These models are then used to synthesize controllers that are robust to errors in perception. If the resulting policy is not safe, we gather new counterexamples. By repeating the process, we eventually find a controller which can keep the system safe even when there is a perception failure. We demonstrate our framework on two scenarios in simulation, namely lane keeping and automatic braking, and show that it generates controllers that are safe, as well as a simpler model of a deep neural network-based perception system that can provide meaningful insight into operations of the perception system.
\end{abstract}

\section{Introduction}


Safety critical autonomous systems like autonomous vehicles (AVs), Unmanned Aerial Vehicles (UAVs) and surgical robotics are  increasingly relying on learning-based \textit{perception} modules for sensing the world around them. Hence, their safety is dependent on the accuracy of these \textit{perception} modules, which increasingly are Machine Learning (ML)-based. While such perception modules have shown outstanding performance, they are hard to analyze and still prone to errors that can lead to unsafe outcomes \cite{ubercrash}. For safe closed-loop operation of systems that use these them, it is important to design controllers which are robust to the errors in such perception modules. In this work, we develop a framework that enables us to synthesize simple {\em surrogate models} of the perception module and use them to design controllers that are robust in the sense that the system satisfies a safety property specified in temporal logic. In particular, we focus on a simulation-based approach for autonomous systems, especially AVs that rely on ML-based perception modules to sense and understand the environment for closed loop decision making and control. 

 \begin{figure}[t]
 \vspace{0.15cm}
        \centering
        \includegraphics[width=0.45\textwidth]{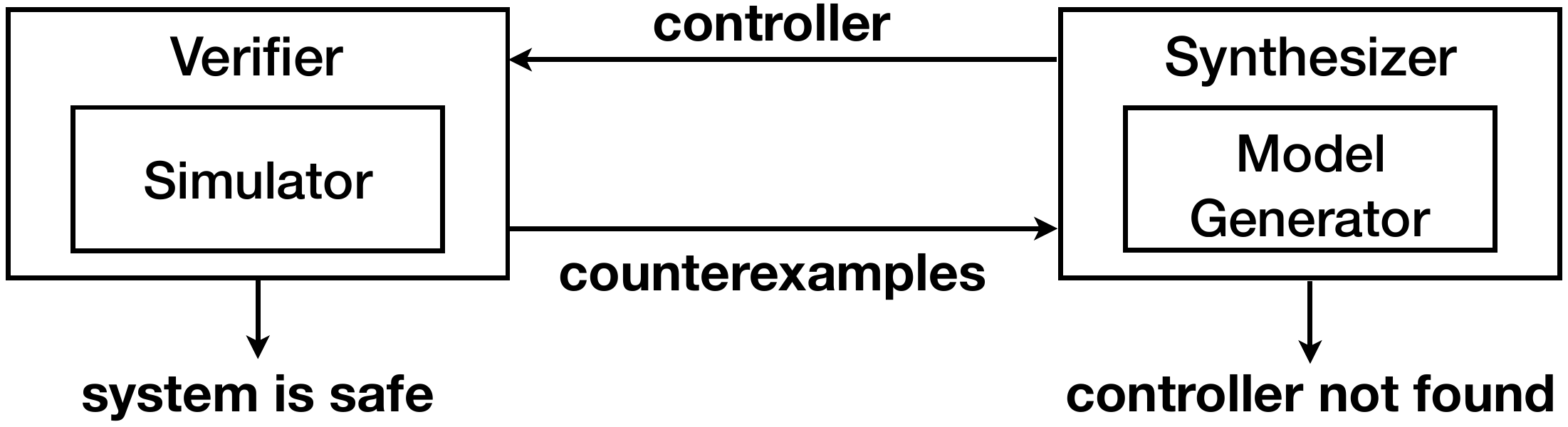}
        \vspace{-5pt}
        \caption{\small{Overview of our framework.}} \label{fig:overview}
        \vspace{-10pt}
\end{figure}

\begin{figure}[b]
        \centering
        \vspace{-15pt}
        \includegraphics[width=0.3\textwidth]{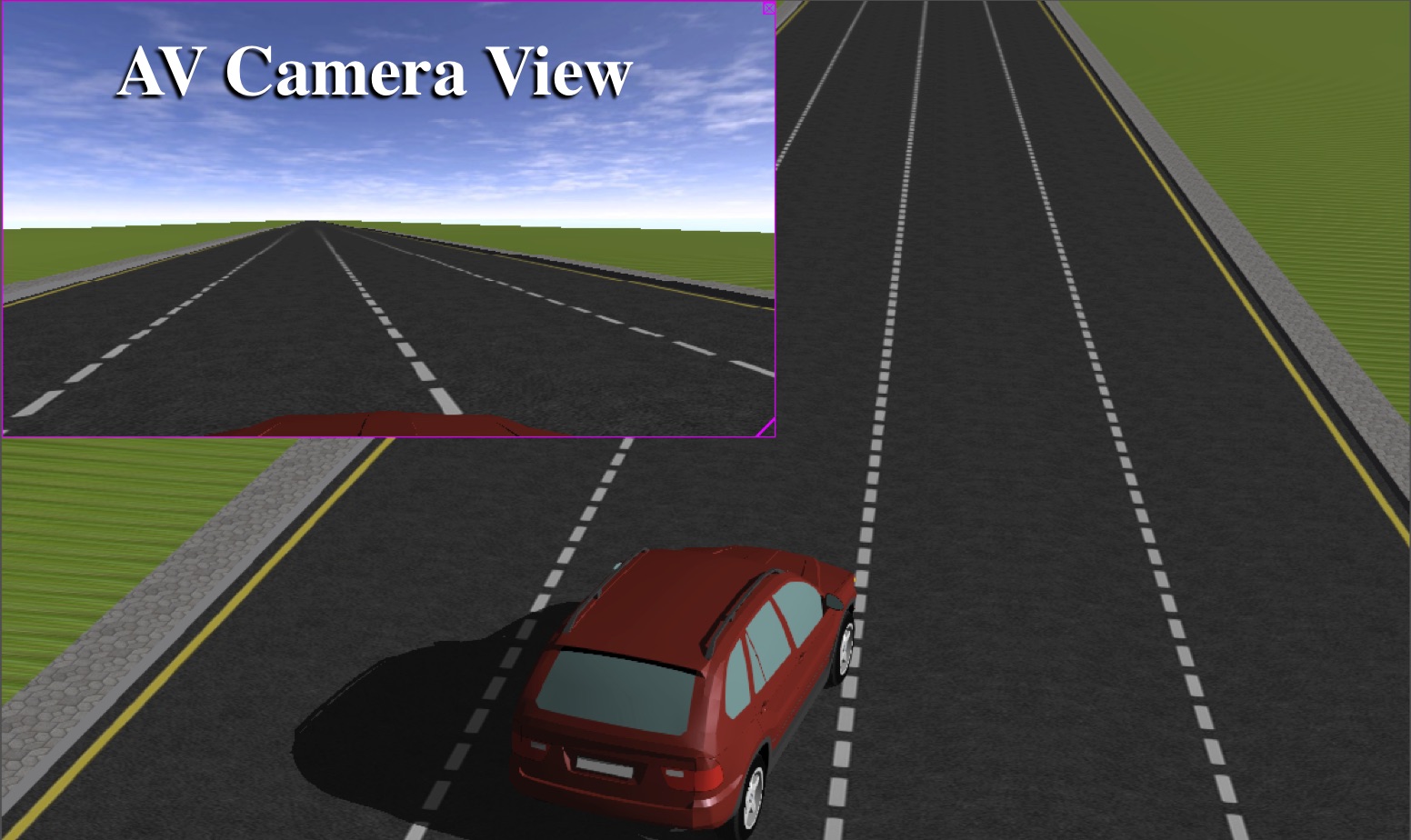}
        \vspace{-5pt}
        \caption{\small{Webots~\texttrademark \ for design and analysis of AVs.}} \label{fig:webots}
        \vspace{-10pt}
\end{figure}

\noindent \textbf{Overview of our approach:}
Synthesizing robust controllers without an explicit white-box analysis of ML-based perception modules can be very expensive, at best, and impossible, at worst, due 
to the complexity of these modules for for control design. 
Our approach overcomes this by using the paradigm of {\em oracle-guided inductive synthesis}~\cite{seshia-arxiv16,jha-acta17} to synthesize a perception model 
using counterexamples that are generated iteratively during the control design process.
We begin with an arbitrary controller, a simulatable model, and a safety property in a suitable temporal logic. A \textit{verifier} performs temporal-logic falsification (e.g. \cite{STaliro,verifai}), simulating the controller in different scenarios until it finds one where the behavior is unsafe. Data from such runs are marked as counterexamples. These counterexamples are used by the \textit{synthesizer} to learn a perception model which is then used to design a new candidate controller. This process repeats, till either a controller is found with which the safety property cannot be falsified, or no such controller can be found. Figure \ref{fig:overview} shows an overview of this framework. Note that our approach jointly synthesizes {\it both} the controller and the surrogate perception model. We use state-of-the-art simulators (such as Webots\texttrademark, see Figure~\ref{fig:webots}) for our experiments including to generate data which is used to build the simplified models of perception modules.

%
While modeling ML-based perception systems accurately is a challenge, we approach this problem by synthesizing simple surrogate model.
Such models map each output $\vy$ generated by the actual perception component to a set of possible values that could correspond to $\vy$. While the form of this surrogate perception model needs domain expertise, its synthesis is completed using both unsupervised techniques such as clustering and the counterexample-guided synthesis process. 
This simpler perception model is used for control synthesis, yielding a fully automated synthesis procedure.  
Note that the surrogate model is not meant to replace the actual complex perception module, it only approximates the subset of the domain where the controller is not robust to the complex module's behavior.

\begin{example}[Lane keeping]
\label{ex:running}
Consider an Autonomous Vehicle (AV) maintaining its lane in the simulator. The AV's perception unit has (a) a camera that is mounted in front of the AV, to estimate its position with respect to the center of the lane it is following, and (b) a compass along with an imprecise GPS which can estimate orientation of the AV w.r.t. the road. A regression-based perception module takes uses the sensor data to detect lane boundaries and estimate the AVs deviation from the lane center. These estimates are sent to a feedback controller which tries to minimizes distance between the car and the lane center by steering the AV. Figure \ref{fig:webots} illustrates this in the Webots simulator.
\end{example}

We will use the lane keeping task (described above) as a running example to explain the main components of our framework, i.e. the synthesis of the surrogate model for the perception module and the lane keeping controller; and how the verifier is used in the iterative process. 

%

{\bf Contributions:} The main contributions of this work are: 
\begin{itemize}
\item A novel counterexample-guided method to synthesize controllers robust to perception errors;
\item A data-driven approach for inference of simple models of complex perception modules, including ML-based perception, and
\item Two case studies from the domain of autonomous driving: (i) lane-keeping where a perception module uses classical vision-based algorithms, and (ii) automatic braking with a Neural Network (NN)-based perception module; demonstrating that our framework is general enough for both ML-based and non-ML-based perception modules.
\end{itemize}

Through the simulation studies, we see how our framework can synthesize {\em both} a robust controller and surrogate perception models that provides insight into how the otherwise complicated perception module performs, e.g. regions where the NN-based perception module can fail, as well as ranges around the true values where its output could lie and how the synthesized controller takes them into account.



{\em Paper Outline:} 
We briefly discuss the state-of-the art in Section \ref{sec:related}. Section~ \ref{sec:prelimandprob} formally sets up the problem we aim to solve in this work. Section~\ref{sec:framework} presents our framework for counter example-guided synthesis of controllers and surrogate perception models, with Section~\ref{sec:modellearning} outlining how we learn the simpler surrogate models for otherwise complex perception modules. Two simulation studies in Section~\ref{sec:simulations} show the applicability of our method, especially to problems related to perception-based control of autonomous vehicles. We conclude with a short discussion in Section~\ref{sec:conclusions}.

\section{Related work}
\label{sec:related}

While advances in Machine Learning (ML)-based perception have led to improved performance of autonomous systems, the proliferation of literature on adversarial attacks (see, e.g.~\cite{GoodfellowMP18}) and verification of ML-based cyber-physical systems (e.g.~\cite{dreossi-nfm17,DJS-cav18,fremont-itsc20}) shows us that state of the art ML-based perception systems are still prone to errors. Closed-loop control design for autonomous vehicles usually involves making assumptions on the perception system and controller that allow both components to be studied independently. One of the primary focuses of perception design is improving local robustness of the perception systems~\cite{li2015,kim2014,jin2015robust,wong2017provable}. Work on designing controllers robust to errors in the perception module \cite{richardsetal05rmp,pant2020tcst} typically model those errors as worst-case disturbances. 

Our framework allows for capturing both a richer class of state-dependent estimation errors and specifications for the behaviors that the closed-loop system should satisfy. It iteratively uses a counterexample-guided approach to build a simple surrogate model of the perception errors and a controller robust to these errors. This stands in contrast with approaches such as counterexample-guided data augmentation~\cite{dreossi2018counterexample} or counterexample-guided environment modeling~\cite{chen2020counter} which directly seeks to improve the accuracy of perception modules or of specifcations for environment behaviors. Instead of improving perception modules in isolation, we use counterexamples to directly improve the controller robustness. Our experience is that this approach leads to improvement in the overall system safety while using far less data. We build upon existing work on automatically finding counterexamples through simulation-driven falsification (e.g.,~\cite{verifai}). For learning the surrogate perception model, we use a simple clustering method. Similar methods has been used for analysis of perception components in~\cite{Pasareanu2019}, but not in the context of robust control design. 

A similar notion of working with simpler representations of complex systems, in the form of abstraction of dynamical system has been studied rigorously for correct-by-construction control design~\cite{tabuada2009verification}. Abstraction of learning-enabled systems such as neural network has been studied~\cite{pulina2010abstraction,gehr-ieeesp18} but not for the purpose of controller synthesis. Unlike these approaches, we do not aim to obtain guaranteed over-approximate abstractions, but instead obtain surrogate models that can be used for control synthesis.

\section{Preliminaries and Problem Statement}
\label{sec:prelimandprob}
%

\subsection{Problem setup and notations}
\label{sec:prob_setup}
We are interested in synthesizing closed-loop controllers which are robust to perception errors for autonomous agents. In our setup, the autonomous agent (ego) interacts with the external environment in a simulator\footnote{While our framework is applicable to non-deterministic simulators, for simplicity here we consider deterministic simulators}.

\begin{definition}[Simulator]
\label{def:simulator}
A simulator is a tuple $\sys(f_S, h_S, X_S^0)$ where $f_S:X_S\times U \rightarrow X_S$ defines the transition function or the dynamics, and $h_S:X_S \rightarrow Y$ is the output or perception function. $X_{S}^0 \subseteq X_S$ is the set of possible initial states.
\end{definition}

The perception module processes the simulated sensor data, e.g. the camera data in Webots for lane keeping (example~\ref{ex:running}), and provides signals $\vy \in Y$ (e.g. estimates of deviation from center of lane) for the controller $\cont$ to act on. The controller applies a control signal $\vu \in U$ to the simulator, and this results in an evolution of the simulator state\footnote{The controller does not have direct access to the state of the simulator, and relies on the perception function $h_S$ for feedback} $\vx_s \in X_{S}$. For the lane keeping task (example~\ref{ex:running}), the perception function $h_S$ represents the lane detector and regression-based estimate of the AV's deviation from the center of lane. 

\begin{remark}
For our simulation-based setup, the output function $h_S$ is composition of two parts. The first is the renderer which given a state of the simulator $\vx_s$ produces the associated sensor data e.g., images and point clouds. These sensor readings are sent to the perception modules which provide the measurement $\vy$ required by the controller $\cont$. We consider the composition of the renderer and perception modules as the perception function.
\end{remark}

\begin{definition}[Controller]
A controller, or the control policy with parameters $\vp \in P$, is deterministic and defined as $\cont:P \times Y \rightarrow U$.
\end{definition}
Trajectories, or traces of the closed-loop system formed by the simulator (including the perception module) and controller (figure~\ref{fig:closed-loop}) are defined by $\xi_S:(\vx_S(\cdot), \vy(\cdot), \vu(\cdot))$, wherein,
for a given $\vp \in P$, initial state $\vx_S(0) \in X_{S}^0$, and all (discrete) times ($\forall i \in \mathbb{N}_{\geq 0}$):

\vspace{-10pt}
{\small
\begin{align*}
	\vx_S(i+1) & = f_S(\vu(i), \vx_S(i)) \\
	\vy(i) & = h_S(\vx_S(i)) \\
	\vu(i) &= \cont(\vp, \vy(i))\,.
\end{align*}}
\vspace{-10pt}

\textit{Example 1 (Lane keeping: The closed loop system)} In this example, simulator state $\vx_S$ includes (i) state of the AV ---position, orientation and speed of the AV, and (ii) potentially time-varying environment parameters such as time of the day, and AV's target lane on the map. 
Given a state of the simulator $\vx_S$, the simulator renders the image seen by the AV's camera. The perception module processes this image to estimate the distance to the lane center $d$. Similarly GPS and compass readings are used to measure the AV's relative orientation to the lane. These form the processed information ($\vy$) which is fed to the feedback controller to compute the steering ($\vu$) required to stay in the lane. We consider a linear control law of the form $\vu= \cont(\vp,\vy)=\vp^T\vy$. More details in section~\ref{sec:simulations}. 

The controller is tasked with ensuring that all traces of the closed-loop system satisfy a \emph{safety specification}, defined over a finite time horizon $H$ as:

\begin{definition}[Specification]
Given a time-varying safe set $\mathfrak{S}_S: \integers \rightarrow 2^{X_S}$ and a trace $\xi_S$, the trace satisfies a safety specification $\spec_S$, i.e. $\xi_S \models \spec_S$, when $\vx_S(i) \in \mathfrak{S}_S(i)\, \forall i \in \{0,H\}$.
\end{definition}

Finally, we define counterexamples for a given specification (and system).
\begin{definition}[Counterexample]
\label{def:counterexamples}
Given system $\sys$ and controller $\pi$, traces of the closed loop system $\xi_S$ such that $\xi_S \not\models \spec_S$, i.e. traces that do not satisfy the specification $\spec_S$, are called counterexamples for $\spec_S$.
\end{definition}

\subsection{Problem Statement}
\label{sec:problemstatement}

In this paper, we are interested in solving the following problem:

\begin{prob}[Control Synthesis]~\label{prob:main} Given a simulator $\sys$, a specification $\spec_S$, and a  control policy $\cont$ parametrized by $\vp$, find $\vp$ such that all traces $\xi_S$ of the closed loop are safe, i.e. $\xi_S \models \spec_S$. 
\end{prob}

\textbf{Challenge in solving this problem:} In its current form, solving this problem is usually intractable as the simulator $\sys$, which includes the perception component as well as state transition dynamics (e.g. for the AV and the environment), is too complex to be used for controller synthesis. 

\textbf{Simple surrogate model:} In order to overcome this, we wish to have a simple but non-deterministic surrogate model for the simulator (and the perception module). We emphasize that we are not interested in general purpose models, but models that are tailored specifically for solving the control synthesis problem defined above. Let $\vx_M \in X_M$ be the state of the surrogate model. The inputs to the model are the same as the input signals to the simulator $\vu \in U$. Since the surrogate model is meant to be less complex than the simulator, it aims to generate measurement signals that for a given input, would over-approximate the output of the simulator. This is formalized below:

\begin{definition}[Surrogate Model]~\label{def:model}
	A surrogate model $\model$ of the simulator $\sys$ is a tuple $\model(f_M,h_M,X_{M}^0)$, where $f_M:X_M\times U  \rightarrow 2^{X_M}$, $h_M:X_M \rightarrow 2^Y$, and $X_{M}^0$ is the initial set of states.
\end{definition}

\begin{figure}[t!]
 \vspace{0.15cm}
    \centering
    \begin{subfigure}[t]{0.2\textwidth}
        \centering
        \includegraphics[width=0.8\textwidth, trim={.2cm .5cm .2cm .2cm}]{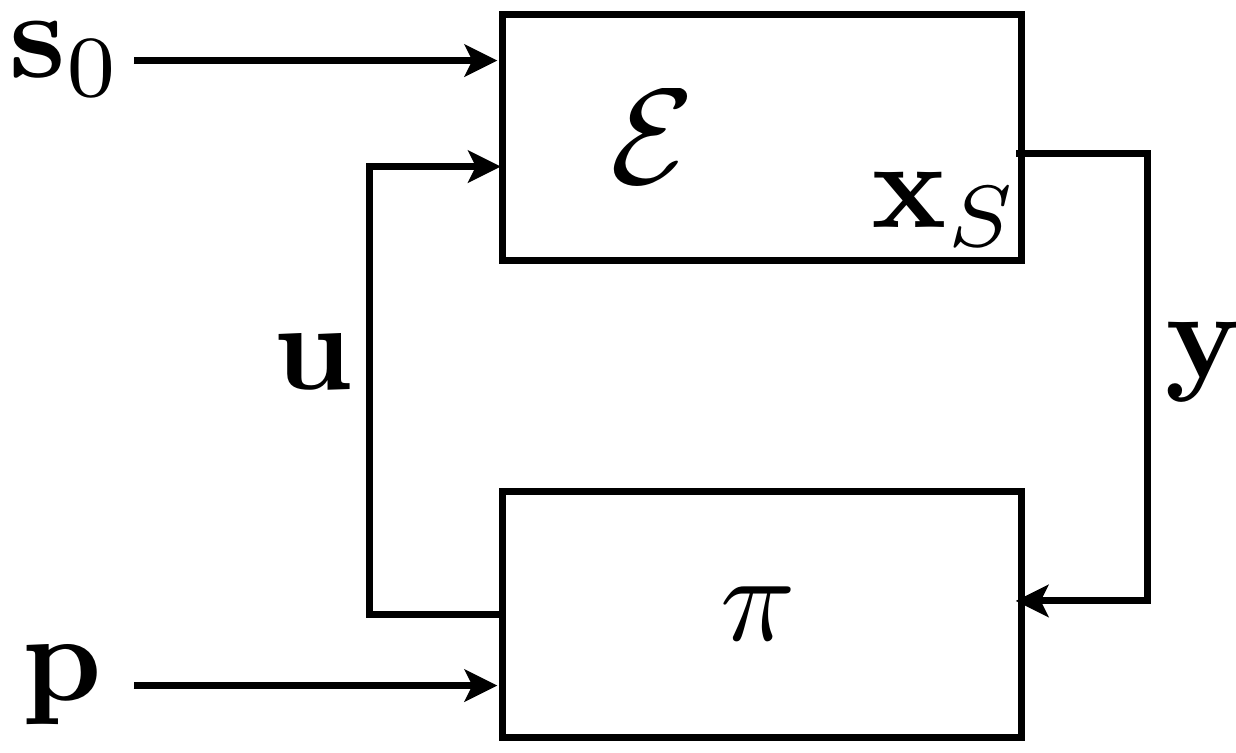}
        \caption{simulator in the loop} \label{fig:closed-loop} 
    \end{subfigure}%
    ~ ~ ~
    \begin{subfigure}[t]{0.2\textwidth}
        \centering
        \includegraphics[width=0.8\textwidth, trim={.2cm .5cm .2cm .2cm}]{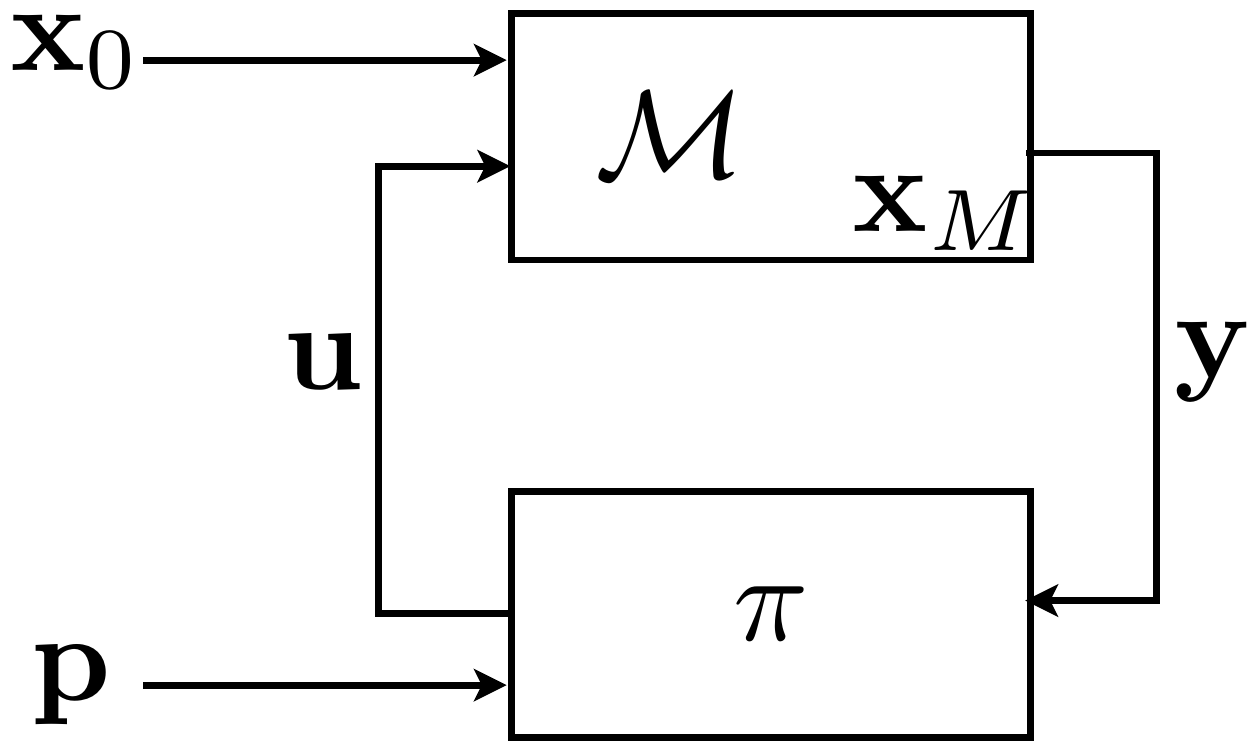}
         \caption{model in the loop} \label{fig:model}
    \end{subfigure}
    \vspace{-5pt}
    \caption{\small{Schematic view of closed-loop systems.}}
    \vspace{-10pt}
\end{figure}

\noindent\textbf{Note:} Designing this surrogate model requires a domain expert to provide an initial form for $f_M$ and $h_M$. The next section covers how this surrogate model is refined through our counter-example guided framework.

For the model , we define a transition \emph{relation} ($f_M$) which models the dynamics, and an output or perception \emph{relation} ($h_M$) which captures imperfect perception by mapping a state of the surrogate model to \text{possibly multiple} output values. A trace of the closed-loop system with the surrogate model (Fig.~\ref{fig:model}) $\xi_M$, is defined similar to $\xi_S$, by replacing $\vx_S(\cdot)$ with $\vx_M(\cdot)$. For a given $\vx_M(0) \in X_{M}^0$
\[
\vx_M(i+1) \ \in \ f_M(\vx_M(i), \vu(i)) \ \ , \ \ \vy(i) \ \in \ h_M(\vx_M(i)) \,.
\]


To use the surrogate model for solving the control synthesis problem (problem~\ref{prob:main}), we assume the following two properties of the surrogate model that allows us to relate it to the simulator:

\begin{assu}
\label{ass:stateofM}
The state $\vx_M \in X_M$ of the surrogate model includes the relevant information required for synthesizing a controller $\pi$ for $\sys$.
\end{assu}
This ensures that model is indeed able a good surrogate for the true perception system for the controller and we do not lose any relevant information for controlling the system. In our running example, we would like the model to provide us the positional deviation from the lane center and heading deviation for synthesizing the controller. If the heading (or positional deviation) was not available, we would not know which direction to steer (or how much to steer).

\begin{assu}
\label{ass:alpha}
We assume the model state $\vx_M$ can be efficiently computed from a simulator state $\vx_S$, i.e., there exists a function $\alpha$ that can map every state $\vx_S$ to its corresponding model state $\vx_M$ ($\alpha : X_S \rightarrow X_M$).
\end{assu}

\noindent \textit{Example 1: Surrogate model for lane keeping.} For the running example, the surrogate model $\model$ of the AV would require a state $\vx_M$ that captures (a) relative orientation of the AV w.r.t. the target lane, (b) relative deviation from the center of the target lane, (c) speed of AV. $\vx_M$ only stores the relative position (to the lane center) instead of the AVs position in a global coordinate frame as $\vx_S$ does. $\vx_M$ ignores other information in $\vx_S$ including the target lane on the map since it is not relevant for synthesizing the controller. 

\begin{definition}[Specification on surrogate model]
For a finite-horizon safety specification $\spec_S$ on the simulator $\sys$,  we define a corresponding specification $\spec_M$ on the surrogate model $\model$ s.t. $\xi_M \models \spec_M \Rightarrow \vx_{M}(i) \in \mathfrak{S}_M(i) \forall i \in \{0,H\}$, where $\mathfrak{S}_M(i)$ are time varying safety sets.

\end{definition}

Before describing how we use the surrogate model to solve problem~\ref{prob:main}, we assume the following:

\begin{assu}
\label{assu:domain}
In this work, we assume domain $X_M$, along with $\alpha$, $X_{M}^0$, and $\spec_M$ are provided by an expert. To complete our model, we need to define $f_M$ and $h_M$.
As we focus on modeling perception units in this work, for simplicity we assume a transition relation $f_M$ can mimic $f_S$ is provided by an expert using a system identification process: $\vx_S' = f_S(\vx_S, \vu) \implies  \alpha(\vx_S') \in f_M(\alpha(\vx_S), \vu)$

\end{assu}

\noindent \textbf{Note:} This ensures any transition that is produced by the simulator is producible by the model as well. Hence, the only missing part of the model is $h_M$, and we wish to learn that systematically. With this, we would like to solve the following, potentially simpler, problem:

\begin{prob}[Synthesizing a Surrogate Model \& Controller]
\label{prob:model} 
Given the simulator $\sys$, specification $\spec_S$ and a specification for the the surrogate model $\spec_M$, we want to synthesize: a surrogate model $\model$ and the controller $\pi$ (parameters $\vp$) such that:

\emph{All closed loop traces of $\model$ are safe $\xi_M \models \spec_M \Rightarrow$ all traces of the $\sys$ are also safe $\xi_S \models \spec_S$.}
\end{prob}

Here, we are using the simpler surrogate model to synthesize a controller to satisfy $\spec_M$, such that the controller also satisfies the specification for the simulator $\spec_S$. We show through the simulation examples that this problem is tractable, as opposed to problem~\ref{prob:main}. We describe how our framework solves problem~\ref{prob:model} in the next section.






\section{The Synthesis Framework}
\label{sec:framework}


\begin{figure}[t]
 \vspace{0.15cm}
        \centering
        \includegraphics[width=0.44\textwidth]{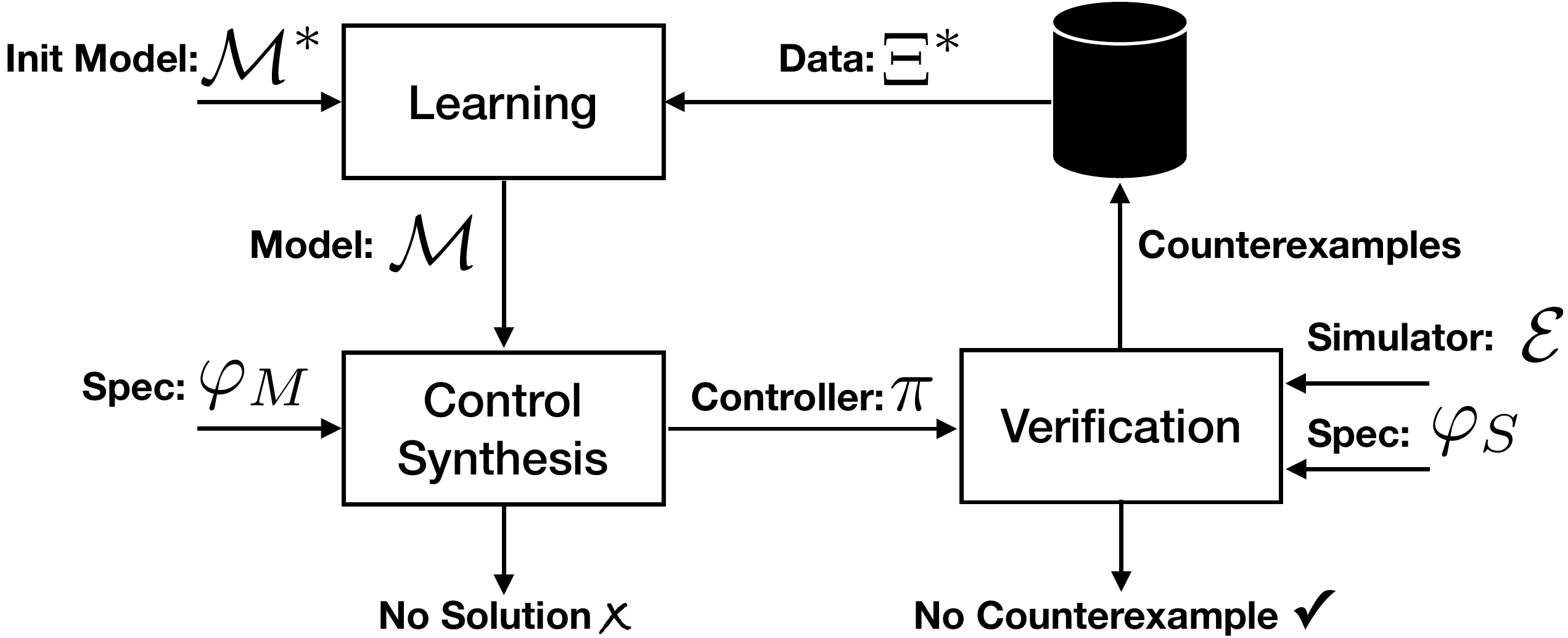}
        \vspace{-5pt}
        \caption{\small{The Counterexample-guided framework for synthesizing a controller through learning surrogate models for the complex perception module.}} \label{fig:learning-loop}
        \vspace{-20pt}
\end{figure}

We solve the problem~\ref{prob:model} by iterating over the surrogate models, and synthesizing controllers for each of them. We use the counterexamples generated by the falsification engine (for falsifying $\spec_S$) for the iterative refinement of the surrogate model. This process is detailed in this section. 

Our framework has three main components: (a) a control synthesizer, (b) a system verifier, and (c) a surrogate model generator. 
Fig.~\ref{fig:learning-loop} shows how the components interact. We start with an initial surrogate model\footnote{Given by the domain expert as in assumption~\ref{assu:domain}} $\model^*$; and in each iteration, we first solve a control synthesis problem using the simpler surrogate model.

\noindent\textbf{1. Control Synthesis:} 
We use Counter Example Guided Inductive Synthesis (CEGIS)~\cite{solar2006combinatorial} to synthesize parameters $\vp$ for the controller $\pi$ such that closed loop system ($\model$ and $\pi$) satisfies $\spec_M$. Briefly, this method iteratively synthesizes a candidate policy and tests it over the surrogate model. It uses gradient-based search with random initialization to search for parameters $\vp$. We omit the details of this procedure for brevity. In general, our framework can incorporate any control synthesis procedure that uses gradient based techniques to generate the controller parameters $\vp$.


\noindent\textbf{2. System Verifier:} 
Once a policy $\cont$ is obtained, we need to verify if it is safe for the simulator as well. For this purpose, we use a falsification procedure (we refer the interested reader to \cite{verifai}, \cite{STaliro} for more details) to find counterexamples (see definition \ref{def:counterexamples}) for $\spec_S$. As active sampling-based methods have been shown to be effective in falsifying black-box closed-loop systems~\cite{ghosh2018verifying}, we employ VerifAI, a recently developed toolbox for analysis of AI-based systems, to implement this sub-procedure~\cite{verifai}. Specifically, we use Bayesian optimization for each safety property to find counterexamples of $\spec_S$. We collect these traces and create a dataset. \vspace{-10pt}

\begin{equation}
\vspace{-5pt}
\label{eq:counterset}
\Xi^* \, \text{s.t.} \, \xi_S \not\models \spec_S\, \forall \, \xi_S \in \Xi^*
\end{equation}

If the verifier can not find a counterexample for the synthesized controller $\cont$ within a fixed number of iterations, we declare success. Or else, we use the counterexamples to refine (improve) our surrogate model (Section~\ref{sec:modellearning}).

\noindent\textbf{3. Surrogate Model Generator:} 
In each iteration, we use the generated counterexamples to improve the model. We do so by learning the output relation $h_M$ for a new surrogate model $\model$ from the data set $\Xi^*$ (counterexamples) that satisfies:\vspace{-10pt}

\begin{equation}\label{eq:perception-simulation-relation}
\vspace{-5pt}
\vy(i) \in h_M(\alpha(\vx_S(i)))\, \forall i \in\{0,\dotsc,H\}
\end{equation}

This constraint allows us to argue that as the dataset $\Xi^*$ enlarges and covers more behaviors, it is more likely for a model capable of mimicking the data, to be able to also mimic the simulator, and the likelihood of this approaches $1$ in the limit under some mild assumptions \cite{ghosh2018verifying}. However, it is not practical to gather a huge amount of data covering all possible behaviors. We argue that we need carefully selected data points, where the selection criteria should be determined during the control design process.
Inspired by recent work on counterexample-guided data augmentation~\cite{dreossi2018counterexample}, the key idea here is to aggregate data iteratively in a counterexample-guided loop with the counterexamples provided by the falsifier. At each iteration we determine what types of data are required to improve the model. This iterative data collection allows us to learn a model which is carefully crafted for control synthesis purposes. Details of this follow. 


\subsection{Counterexample-Guided Synthesis of the surrogate perception model and control}
\label{sec:thescheme}
The framework results in an iterative procedure, with assumption \ref{assu:domain} as a starting point, where at each iteration we update the surrogate model by gathering more data. Algorithm \ref{alg:CEG}  outlines the process, and the details of how the surrogate model is learnt from counterexamples. 




\begin{algorithm}[]
\vspace{0pt}
{\small
\SetAlgoLined
\KwResult{Parameters $\vp$ for controller $\pi$, surrogate perception model $\model$}
 \textbf{Initialization:} Simulator $\sys$, initial guess for $\model$ and $\vp$, specifications $\spec_S$ and $\spec_M$, $\Xi^*=\{\}$, Iterating=1\; 
 \While{Iterating}{
  \textit{Controller Synthesis:} Use $\model$ to synthesize $\pi$ to satisfy $\spec_M$\;
  \eIf{Synthesis succeeds, i.e. all $\xi_M$ s.t. $\xi_M \models \spec_M$}{
   Update $\vp$ to output of synthesis procedure\;
   }{
Declare \textbf{failure} and set Iterating=0\;
  }
  \textit{Verification:} Use simulator $\sys$ and falsifier to find all possible counterexamples $\xi_S$ s.t. $\xi_S \not \models \spec_S$\;	  
  \eIf{Counterexamples $\xi_S$ found}{
  Add all counterexamples to $\Xi^*$\;
  \textit{Learn surrogate model:} Use counterexamples in $\Xi^*$ to learn a new surrogate model $\model$\;
  \If{New surrogate model same as previous one}{Declare \textbf{failure} and set Iterating=0\;}
  
  }{
  Declare \textbf{success} and set Iterating=0\;
  }  
 } 
 \caption{\small{Counterexample-guided synthesis of perception model and controller}}
 \label{alg:CEG}}
\end{algorithm}

\subsection{Surrogate Model Generation}
\label{sec:modellearning}
As covered in assumption \ref{assu:domain}, the transition relations $f_M$ for the surrogate model $\model$ are provided by a domain expert, possibly derived from laws of physics with simple models for uncertainties~\cite{ljung1999system}. This leaves us with the problem of learning the output relation $h_M$, which is a simplified surrogate model for the perception module. In particular, perception components are hard to decompose into smaller systems and reason over in a compositional way. Moreover, instead of laws of physics, perception modules are generally artifacts of machine learning or complex vision-based algorithms. A natural way to define $h_M$ then is to use an initial guess $h^*_M$ with an addition of some errors. We model perception errors with state dependent error functions, whose outputs are independent of each other:
\begin{equation}
\label{eq:errormodel}
y_i \in h_{M,i}(\vx_M) = \{h^*_{M,i}(\vx_M)\} \oplus E_i(\vx_M)
\end{equation}

where $E_i : X_M \rightarrow 2^{\reals}$ and $\oplus$ represents a Minkowski sum (set addition). This error model assumes that error of measuring $i^{th}$ component of $\vy$ does not directly depend on other measurement, but depends only on the state. 

\begin{figure}[t]
 \vspace{0.15cm}
        \centering
        \includegraphics[width=0.42\textwidth,trim={.1cm .3cm .1cm .4cm}]{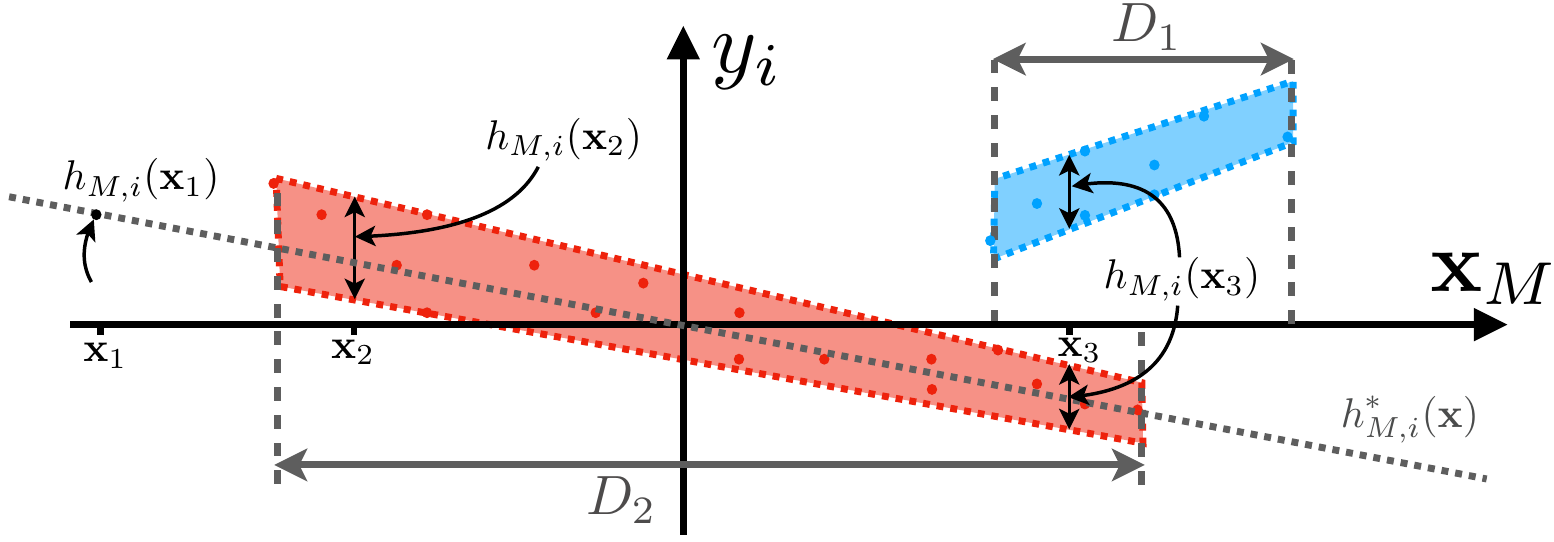}
        \vspace{-5pt}
        \caption{{\footnotesize Modeling the perception relation \eqref{eq:errormodel} for output $y_i$ in the surrogate model, $h_{M,i}$, with clustering. The figure depicts two clusters (in red and blue) for the error $E_i$ added to the initial guess $h_M^*$, which is a linear model here. For a given $x_M$ that lies in a cluster, this perception relation gives a range of possible outputs, e.g. $h_{M,i}(x_2)$.  For a state that lies in multiple clusters, e.g. $x_M = x_3$, the output could be in multiple ranges.}} \label{fig:clustering}
        \vspace{-10pt}
\end{figure}

Note that we rely on the designer to choose a low-dimensional state space that is representative enough of the perception model. For example, if the perception model is sensitive to the color of an environment car we need to include it in the state space of the surrogate model.

\noindent\textbf{Unsupervised learning of the error model:}
Next, we investigate an unsupervised learning technique for modeling $E_i$.
Our approach is to cluster the datapoints and build a local model for each cluster. 
Recall that we use counterexamples (Eq.~\eqref{eq:counterset}) to learn the model,
but these may be ad-hoc because (i) errors in learning-enabled components can be ad-hoc, and (ii) counterexamples reveal errors in the entire closed-loop system. 
Clustering helps to better model errors for regions in which perception may not be accurate.

\noindent\textit{Clustering the counterexamples:} Initially datapoints $(\vx_S, y_i)$ are extracted from $\Xi^*$ and are then projected into $X_M$: $(\vx_M = \alpha(\vx_S), y_i)$. Next, we use standard clustering algorithms such as KMeans or Gaussian Mixture Models to partition the dataset into different clusters~\cite{robert2014machine}. Let $\{(\vx_M^k, y^k_i)\}_{k\in K_j}$ be data points in the $j^{th}$ cluster. 

\noindent\textit{Local error model for each cluster:}
For simplicity we define the domain $D_j \in X_M$ of the $j^{th}$ cluster to be the smallest box containing $\{\vx_M^k\}_{k\in K_j}$ and $R_j(\vx_M): [low_j(\vx_M), up_j(\vx_M)] \subseteq Y_i$ defines the error range for that cluster. Next we use a linear model for $low_j$; i.e. $low_j(\vx_M) = A \vx_M + B$. We solve for $A$ and $B$ using linear programming such that $A \vx_M^k + B \leq y^k_i \ (k\in K_j)$. A similar procedure is used for $up_j$. 

Figure \ref{fig:clustering} shows a visualization of this process for a linear model as the initial guess of $h_M^*(\vx_M)$, and two clusters for the error model. Notice that clustering is performed in domain $X_M \times Y_i$, but domain of each cluster is defined over $X_M$, therefore a given $\vx_M$ can belong to multiple clusters, as depicted in figure \ref{fig:clustering}. Finally, to obtain a surrogate percetion model as in \eqref{eq:errormodel} we define $E_i$ as follows: 

{\small
\vspace{-10pt}
\[
E_i(\vx_M) = \begin{cases}
	\{0\} & \vx_M \not\in \bigcup_{j} D_j \\
	\bigcup_{\{j \ | \vx_M \in D_j\}} R_j(\vx_M) & \mbox{otherwise}\,.
\end{cases}
\]}
If $\vx_M$ does not belong to any cluster, we assume the error is zero, e.g. see $x_M=x_1$ in Fig.~\ref{fig:clustering}. Otherwise, we consider the ranges for all clusters with $\vx_M$ in their domain to guarantee Eq.~\eqref{eq:perception-simulation-relation} holds for $\model$. This is how the surrogate model can provide multiple values or ranges of values for the output of perception for a given state $x_M$ of the surrogate model.

The number of clusters can be chosen dynamically to reduce the model error in the surrogate model. While we would like to minimize the number of clusters, we can increase it if our error is not within bounds.

\section{Simulation studies}
\label{sec:simulations}

We evaluate our framework through two simulation case studies for AVs with learning-based perception modules. While we use Webots~\texttrademark~\cite{webots}, our technique can be used with many other simulators, e.g. CARLA~\cite{Dosovitskiy17}.
We use a CEGIS approach built on top of the falsification engine in VerifAI~\cite{verifai} to synthesize the controller. Off-the-shelf clustering and linear programming methods are used for the learning the surrogate model.


\subsection{Case Study I -- Lane Keeping}

For the first case study, our running example~\ref{ex:running}, we consider the problem of an Autonomous Vehicle (AV) tasked with lane keeping on a straight roads. For the simulations, we assume initially AV deviates $[-0.4, 0.4] m$ from the lane center and its orientation is $[-0.25, 0.25]$ radian relative to lane orientation. The initial speed $[15, 25] kph$.

\noindent\textbf{Simulator:} The AV can accurately estimate its orientation $\theta_{AV}$ using a compass and the orientation of the road $\theta_R$ using a HD map and an imprecise GPS. However, for the deviation from the lane center $d$, it relies on a camera and computer vision-based approach to get an estimate $\hat{d}$ of $d$. These form the output $y$ of the simulator $\sys$. We next describe the perception module, or the output relation $h_S$: The image taken by the camera is processed to detect lane boundaries~\cite{lane_detector} and a regression process is used to learn a model for estimating deviation using detected boundaries. However, this deviation estimation $\hat{d}$ is not always reliable as it involves image processing and machine learning. The physics-based dynamics of the AV and environment in Webots describe the state dynamics $f_S$ for $\sys$. The information used by simulator e.g. AV position, speed, orientation, distance from center of lane etc. form the state $x_S$. 

\noindent\textbf{Controller:} 
Since we do not have any cars or other agents in the environment, the controller simply performs
lane keeping. 
The steering policy $\pi$ is a linear feedback law w.r.t. relative orientation $\theta_\Delta: \theta_{AV} - \theta_R$ and the estimate of deviation $\hat{d}$; $\pi(\vp,y) = p_1 \theta_\Delta + p_2 \hat{d}$.

\noindent\textbf{Specification:} The goal is to keep the deviation bounded, $d \in [-1, 1]$ and get to $\{(\theta_\Delta,d) \ | \ \theta_\Delta \in [-0.1, 0.1], d \in [-0.3, 0.3]\}$ within $4$ seconds. The time-varying safe sets that capture this requirement for the specification $\spec_S$. For the verifier, the requirements considered in these simulations are represented using Metric Temporal Logic (MTL) using the \textit{Always} and \textit{Eventually} operators. This is however beyond the scope of this work  and we refer the reader to \cite{verifai}.

\noindent\textbf{Surrogate model:}
Recall from section \ref{sec:prob_setup} that the model state $\vx_M$ includes (a) deviation $d$, (b) relative orientation $\theta_\Delta$, and (c) speed $v$. The perception output is the measured state ($\vy^t : [\hat{v}, \hat{\theta}_\Delta, \hat{d}]$).
We assume measurements of speed $v$, and orientations $\theta_{AV}$ and $\theta_R$ are accurate, $h_{M,1}(\vx_M) = v, h_{M,2}(\vx_M) = \theta_\Delta$. We only model error in estimating $d$. $h_{M,3}$ defined over $d$, $\theta_\Delta$, and $v$ and provides a range for $\hat{d}$. However, it is easy to see that speed of AV does not affect measurement of $d$ as we rely on a single image to estimate $d$. Thus, for simplicity we model $h_{M,3}$ as a map from $d$ and $\theta_\Delta$ to range of possible $\hat{d}$. Following \eqref{eq:errormodel} this gives us a surrogate perception model of the form $h_M: [\hat{v},\hat{\theta}_\Delta,\hat{d}] = [v,\theta,d] + E(x_M)$. Note again that $d$ is not the actual output that the controller $\pi$ has access to, it is instead a camera-based estimate $\hat{d}$, while $d$ is a state of the simulator (and surrogate model). The initial guess for the surrogate perception model then is simply the linear model $h_M^* = x_M = [v,\theta,d]$. Finally, the specification $\spec_M$ for the surrogate model is the same as $\spec_S$.

\noindent\textbf{Results:} Starting from the surrogate model for perception $h_M^*$ the control synthesis procedure yields parameters $p_1 = -0.5$, $p_2 = -0.8$. However, the verifier could find a set of counterexamples\footnote{These counterexamples are corner in our simulations and did not happen in a uniformly random generated data.} $\Xi^*$ using Bayesian optimization~\cite{ghosh2018verifying}, since the surrogate model is not adequate in its current form. Next, these counterexamples are used to learn a better model for perception and the process repeats as in algorithm \ref{alg:CEG}. After a couple of iterations, the procedure terminates successfully. 

\begin{figure}[t]
 \vspace{0.15cm}
        \centering
        \includegraphics[width=0.48\textwidth, trim={0cm 0.15cm 0cm 0.2cm},clip]{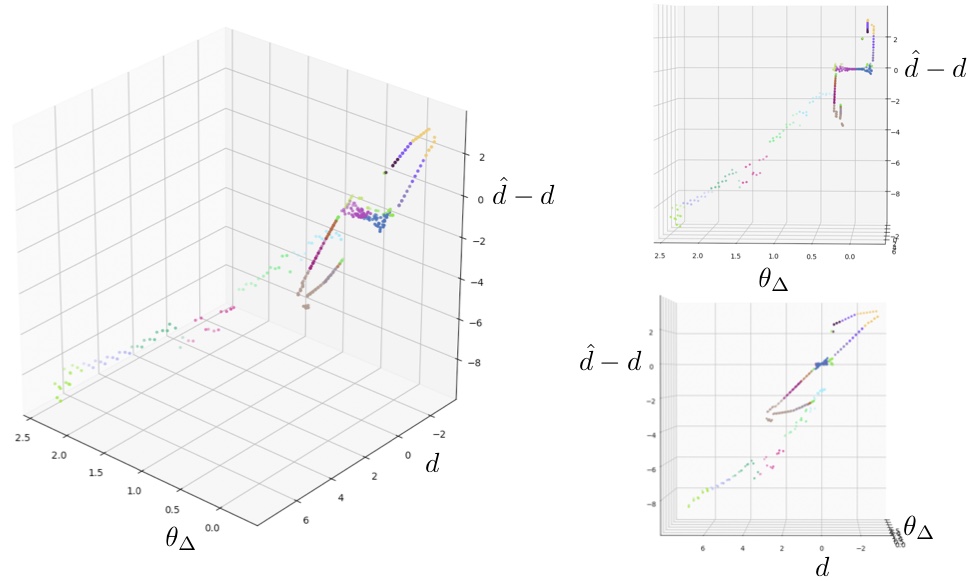}
        \vspace{-20pt}
        \caption{{\small Data extracted from counterexamples for lane-keeping. Points with same colors belong to same clusters.}} \label{fig:clustered-data}
        \vspace{-20pt}
\end{figure}

Set of data points extracted from counterexamples are shown in Fig.~\ref{fig:clustered-data}. The figure shows error on $\hat{d}$ as a function of $d$ and $\theta_\Delta$. As depicted, when $d$ and $\theta_\Delta$ are close to origin, the error is small and the error gets unpredictably large in other places. Using this data extracted from only $11$ counterexample traces we find that parameters $p_1 = -3.93$ and $p_2 = -0.63$ generated for the learned model is robust enough that the verifier can not find counterexamples. 

\noindent\textbf{Insights:} For the obtained controller, intuitively, because the error in measurement of $d$ is relatively large compared to $\theta_\Delta$, the feedback law $p_1 \theta_\Delta + p_2 \ d$ is safe only when $|p_2|$ is relatively smaller than $|p_1|$.  

\looseness = -1

\subsection{Case Study II -- Emergency Braking System}
\begin{figure}[t]
 \vspace{0.15cm}
        \centering
        \includegraphics[width=0.40\textwidth]{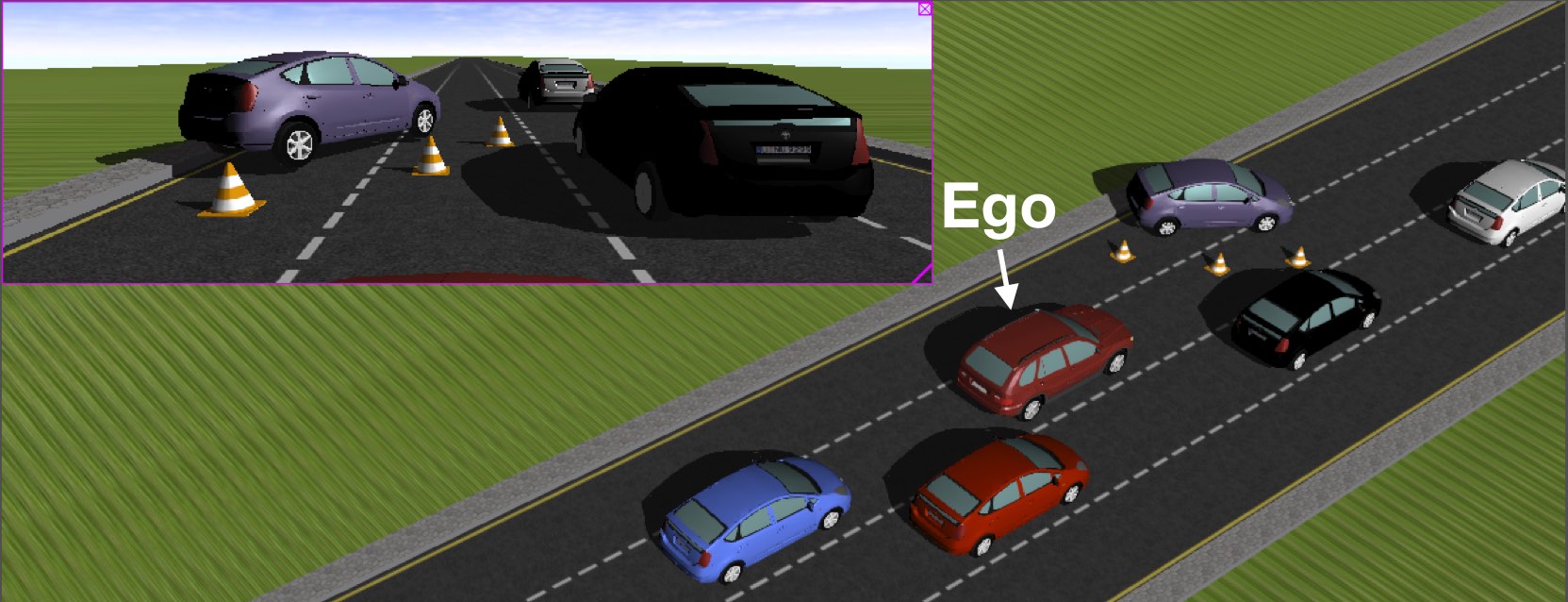}
        \vspace{-5pt}
        \caption{{\small Emergency braking scenario.}} \label{fig:AEB}
        \vspace{-15pt}
\end{figure}

For the second case study, we consider scenarios (in the Webots simulator) in which the AV detects construction cones on the road and brakes to avoid hitting them. In these scenarios two lanes are blocked by a broken car and cones are used to warn drivers. Figure \ref{fig:AEB} shows one such scenario.

\noindent \textbf{Brief description of simulator and surrogate model:}
While the simulation environment can have many variations, including: color of the broken car, orientation of the broken car and speed of the traffic (all environment vehicles have same constant speed), the surrogate model only includes distance to the cones\footnote{In this case study, we use $d$ to denote the distance of the AV to the cones, unlike the previous case study where $d$ was deviation from lane center.} $d$ and speed of the AV in its state $x_M$ and has simple dynamics $f_M$ s.t. $v$: $\dot{d} = -v$, $\dot{v} = u$ where $0 \leq u \leq 2.5$ is the braking force applied by the controller $\pi(\vp,\hat{d})$. The output of the simulator and the surrogate perception module is $\hat{d}$, which is a Neural Network-based estimate of $d$ by using the camera. 

\noindent \textit{The emergency braking system:} A straightforward solution for designing an emergency braking controller is to brake as soon as possible to guarantee safety. While this is the best strategy when the distance to the cones $d$ is small, it reduces passenger comfort and increases chance of accident when there is a car behind the AV. To avoid such behavior we consider a simulated car that moves behind the AV with the same speed at a distance $d_\text{car}$ and has a full knowledge about the cone and brakes in an optimal way. Then, we require the AV to stop while avoiding a crash with the car behind. 

\noindent \textbf{Specification:}
This leads to a specification $\spec_S$ where we require $d \geq \epsilon_1 > 0$, and $d_\text{car} \geq \epsilon_2>0$, where $\epsilon_1, \epsilon_2$ are small positive numbers. The surrogate model however has no information about $d_\text{car}$, and the specification on it $\spec_M$ is to simply ensure that $d \geq \epsilon_1$. The task of ensuring the car behind the AV does not rear-end it is now dependent on the braking profile of the controller $\pi$.

\begin{figure}[t]
        \centering
        \includegraphics[width=0.4\textwidth]{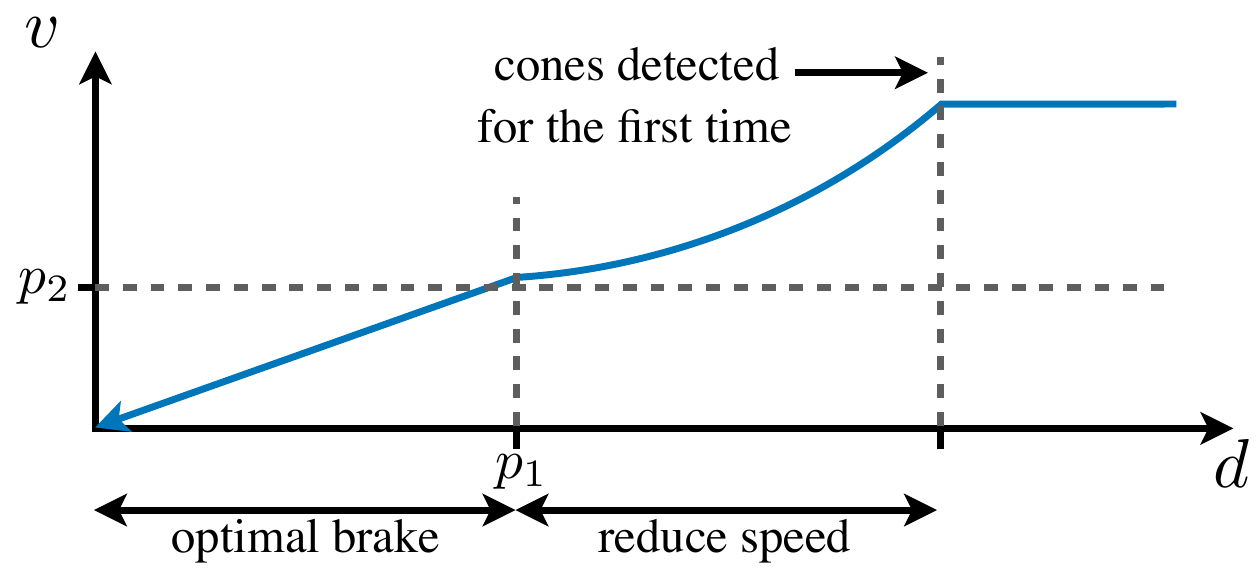}
        \vspace{-5pt}
        \caption{\small{Emergency braking policy.}} \label{fig:brake-policy}
        \vspace{-20pt}
\end{figure}

\noindent \textbf{The emergency braking controller:}
We wish to design a braking control system that uses estimated distance to the cone $\hat{d}$. Knowing the distance and speed of the AV one could use laws of physics to find the minimum force needed for the brake.
In the perception module, the neural network not only detects cones, but also provides a bounding box around detected cones. We wish to use size of these bounding boxes to estimate the distance. However, such a measurement is not reliable especially when the distance is large. To design a safe controller that can satisfy $\spec_S$, we consider the following policy: The controller assumes measured distance $\hat{d}$ is reliable only when $\hat{d} \leq p_1$ and in that case it provides an optimal force: $\frac{v^2}{2\hat{d}}$. However, when $\hat{d} > p_1$, the controller just reduces the speed to $p_2\, m/s$ ($u = p_2 - \hat{v}$) after detecting cones and ignores value of $\hat{d}$ for feedback calculation. Fig.~\ref{fig:brake-policy} shows a qualitative trace of the system.

Recall that $\vx_M$ contains only $d$ and $v$ and output $\vy$ includes estimated value of $d$ ($\hat{d}$) and $v$ ($\hat{v}$). We assume that the speed is known perfectly ($\hat{v} = v$) and since $\hat{d}$ is independent of $v$, $h_{M,2}$ only maps $d$ to range of possible $\hat{d}$ as in \eqref{eq:errormodel}. We also set $\hat{d}$ to infinity if no cone is detected. For using our framework the initial perception model is initialized with a simple linear model as in the previous case study, $h_M^*  = [v,d]$. Finding parameters $p_1$ and $p_2$ for the policy is not a trivial task as these parameters heavily depend on the measurement errors and the dynamics of the agents.

\begin{figure}[t]
 \vspace{0.15cm}
        \centering
        \includegraphics[width=0.49\textwidth,trim={0cm 0.35cm 0 0.2cm}]{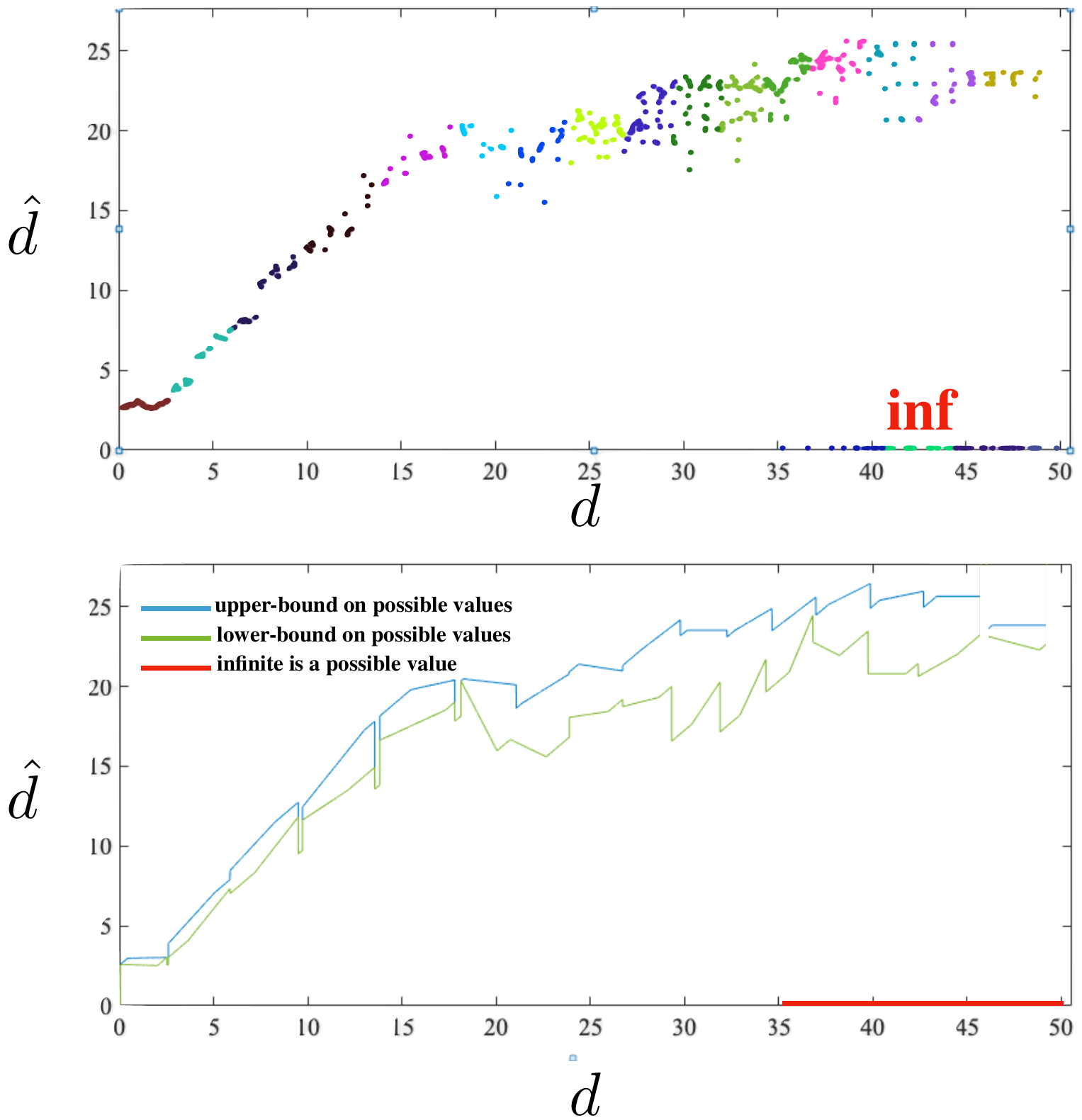}
        \vspace{-10pt}
        \caption{\small{Modeling NN-based perception. Inf means the cone could not be detected. Top figure shows clustered data obtained from counterexamples and bottom figure shows the learned perception relation $h_M$.}} \label{fig:nn-model}
        \vspace{-20pt}
\end{figure}

\noindent \textbf{Results:} 
Initially from $h_M^*$, the control synthesis finds many solutions for the controller parameters $\vp$ that satisfy $\spec_M$ (in closed loop with the surrogate model) and the synthesizer picks $p_1 = 25$ and $p_2 = 7.02$. However, the verifier finds counterexamples, or traces of the simulator that violate $\spec_S$. For these counterexamples, $\hat{d}$ is less than $25$, while the actual distance from the cones $d$ is very larger, and the AV reduces speed quickly crashing with the car behind it, violating $d_\text{car} \geq \epsilon_1$. Using these counterexamples, we update the surrogate perception model $h_M$. In the next iteration only few counterexamples were found, and in all of those cases, it is observed that the color of the broken car\footnote{Which is a simulation parameter that the falsifier can change} is close to the color of the cones. This suggest that the perception unit behaves differently in these cases, causing AV to brake early and collide with the car behind it. Again, by updating the surrogate perception model, the policy synthesizer finds parameters $p_1 = 16.5$ and $p_2 = 7.6$. In other words, the policy uses value of $\hat{d}$ only if $\hat{d} < 16.5$. This strategy allows to safely stop the car such that $\spec_S$ is satisfied such that verifier cannot find any more counterexamples. The final model generated using $40$ counterexamples is shown in Fig.~\ref{fig:nn-model}. Notice that when $d > 35.2$, $\hat{d}$ can be infinity (no cone detected). \looseness = -1

\noindent \textbf{Some insights:} a) The NN-based detector performs poorly when the color of the broken down car is close to the color of the cones, b) despite using a surrogate model and specification $\spec_M$ that ignores the car traveling behind the AV, our framework synthesizes a controller that satisfies the specification $\spec_S$ which in addition to braking before the cones, requires the AV to avoid being rear-ended by the car behind it. This required some domain expertise in the designing the control law $\pi$. c) For the NN-based perception module, our surrogate model finds ranges for the module's output as well as ranges of the true distance to the cone where the perception module fails to detect it, see figure~\ref{fig:nn-model}. The control law does not trust the perception in those ranges\footnote{$p_1 = 16.5$ implies estimate $\hat{d}$ is used by $\pi$ only when $\hat{d}<16.5$}. This clearly shows the simpler surrogate model is not only useful for the control design but also provides useful insight about the otherwise complex perception module. 

\section{Conclusions}
\label{sec:conclusions}

\noindent \textbf{Summary:} In this work we investigated the problem of control synthesis for a closed-loop system with hard to analyze and possibly faulty perception components. Our method iteratively learns models for perception components and then synthesizes controllers for those models. At each iteration we use a falsification-based verifier to put the designed controller under test to prove its safety, or alternatively we find counterexamples which are used to improve the perception models. We demonstrate effectiveness of our method for designing safe controllers for autonomous vehicles which use machine learning-based perception components.\looseness = -1

\noindent \textbf{Limitations and future work:} In order to solve what is otherwise an intractable problem, our approach relies on some simplifying assumptions, including requiring domain expertise in designing the surrogate models. Also, we rely on a falsifier in the verification phase, and for falsifiers, a lack of counterexamples does not always guarantee that the controller always satisfies the given specification. We plan to address these limitations and investigate alternative methods for modeling perception components to make further inroads towards truly guaranteed-safe controller synthesis.

\subsection*{Acknowledgments}
This work was supported in part by NSF grants CNS-1545126 (VeHICaL), CCF-1837132, the DARPA Assured Autonomy program,
by Berkeley Deep Drive, and by Toyota under the iCyPhy center.





\bibliography{refs}
\bibliographystyle{unsrt}
\end{document}